\documentclass[10pt,conference]{IEEEtran}
\usepackage{caption}

\usepackage[T1]{fontenc}
\usepackage[utf8]{inputenc}
\usepackage{cite}
\usepackage{amsmath}
\usepackage{graphicx}
\usepackage{booktabs}
\usepackage{multirow}
\usepackage{xcolor}
\usepackage{relsize}
\usepackage[varqu]{zi4}  
\usepackage{microtype}
\DisableLigatures[-]{encoding=T1,family=tt*}  
\usepackage{fontawesome5}
\usepackage{listings}
\usepackage{enumitem}
\usepackage{url}
\usepackage{orcidlink}
\usepackage{hyperref}
\definecolor{ACMBlue}{RGB}{0,85,162} 
\hypersetup{
    colorlinks=true,
    citecolor=ACMBlue,
    filecolor=black,
    linkcolor=ACMBlue,
    urlcolor=ACMBlue
    }

\newcommand{\code}[1]{\texttt{#1}}

\definecolor{diffadd}{rgb}{0.0,0.4,0.0}
\definecolor{diffdel}{rgb}{0.6,0.0,0.0}
\definecolor{diffcom}{gray}{0.4}
\definecolor{keywordblue}{rgb}{0.13,0.13,1}
\definecolor{stringgray}{rgb}{0.4,0.4,0.4}
\definecolor{darkgreen}{RGB}{0,120,70}

\definecolor{kwc}{RGB}{20,80,175}
\definecolor{strc}{RGB}{176,86,10}
\definecolor{cmtc}{RGB}{40,120,55}
\definecolor{bgc}{RGB}{247,248,250}
\definecolor{frc}{RGB}{200,205,212}
\newcommand{\numLanguages}{187}
\newcommand{\studyLangs}{12}
\newcommand{\corpusFiles}{3{,}000}

\renewcommand{\lstlistingname}{\bfseries Listing}
\makeatletter
\def\fnum@lstlisting{%
  \lstlistingname
  \ifx\lst@@caption\@empty\else~\thelstlisting\normalfont\fi}%
\makeatother

\newcommand{\FrameFiles}{620{,}942}
\newcommand{\SampleProjects}{748}

\newcommand{\lexMedian}{16}
\newcommand{\lexPninety}{25}
\newcommand{\lexMax}{73}
\newcommand{\sharedLoc}{1{,}672}
\newcommand{\FoneCmt}{0.998}
\newcommand{\FoneStr}{0.967}
\newcommand{\UFoneCmtMean}{0.996}
\newcommand{\UFoneStrMean}{0.954}
\newcommand{\FoneCmtPyg}{0.998}
\newcommand{\FoneStrPyg}{0.936}
\newcommand{\CleanDelta}{0.003}
\newcommand{\ParseErrFiles}{370}

\newcommand{\UFiles}{1{,}412}
\newcommand{\UFoneCmt}{0.994}
\newcommand{\UFoneStr}{0.958}

\newcommand{\UProjects}{480}
\newcommand{\URubyProjects}{2}
\newcommand{\URubyFiles}{92}
\newcommand{\UGoProjects}{5}

\newcommand{\ClocN}{1{,}440}
\newcommand{\ClocExact}{0.78}
\newcommand{\ClocCI}{$[0.76,0.80]$}
\newcommand{\ClocMAD}{0}
\newcommand{\ClocRhoMin}{0.96}
\newcommand{\CorrectTableBody}{Python & 0.98 & 0.97 & 3 & 0.38 & Go & 1.00 & 1.00 & 0 & 0.78 \\
Java & 1.00 & 1.00 & 0 & 0.92 & C\# & 1.00 & 0.87 & 0 & 0.90 \\
C & 1.00 & 1.00 & 0 & 0.80 & Scala & 1.00 & 0.97 & 1 & 0.92 \\
C++ & 1.00 & 0.99 & 4 & 0.93 & Ruby & 1.00 & 0.93 & 15 & 0.97 \\
JS & 1.00 & 0.99 & 0 & 0.89 & Bash & 0.99 & 0.92 & 9 & 0.12 \\
TS & 1.00 & 0.97 & 0 & 0.84 & CSS & 1.00 & 0.84 & 1 & 0.87 \\}
\newcommand{\CCmin}{$0.55$ (Ruby)}
\newcommand{\CCmax}{$0.80$ (JavaScript)}
\newcommand{\SCCdiff}{0.06}
\newcommand{\JavaFunShare}{62\%}
\newcommand{\DecRange}{$0.94$ $[0.89,0.97]$ for Java and $0.79$ (Ruby) to $0.98$ (Go) for the other languages}

\newcommand{\WithinTwo}{60--98\%}
\newcommand{\TopOverlap}{55--100\%}
\newcommand{\TopSize}{22--79\%}
\newcommand{\WFlocP}{40}
\newcommand{\WFlocB}{60}
\newcommand{\WFtimeP}{5.0}

\newcommand{\WFtimeB}{7.6}

\newcommand{\WFjavaTone}{6.2\%}
\newcommand{\WFjavaTtwo}{37.4\%}
\newcommand{\WFjavaJscpd}{3.6\%}
\newcommand{\EQfiles}{680}
\newcommand{\EQtimeP}{2.9}
\newcommand{\EQtimePall}{1.6}
\newcommand{\EQtimeB}{8.7}
\newcommand{\EQtimeBall}{9.0}
\newcommand{\EQiqr}{0.3\,s}

\newcommand{\TscanOne}{480}
\newcommand{\Tscan}{2{,}094}
\newcommand{\Tcloc}{880}
\newcommand{\Tclocall}{1{,}866}
\newcommand{\Tscc}{35{,}714}
\newcommand{\LadderAUC}{$1.0$}

\newcommand{\BCBtpNear}{12}
\newcommand{\BCBrNear}{0.012}
\newcommand{\BCBtpJscpd}{2}

\begin{document}

\title{PyReprism: A Shared Lexical Foundation for Polyglot Code Analysis}

\author{
\IEEEauthorblockN{Daniel Ogenrwot~\orcidlink{0000-0002-0133-8164},
Jorge Gonzalo Delgado Cervantes~\orcidlink{0009-0005-5155-8914},
Richard Sserunjogi~\orcidlink{0000-0002-3880-6806},
John Businge~\orcidlink{0000-0003-3206-7085}}
\IEEEauthorblockA{\textit{University of Nevada Las Vegas}\\
Las Vegas, USA\\
\{ogenrwot, delgaj10, sserunjo\}@unlv.nevada.edu, john.businge@unlv.edu}
}
\maketitle

\begin{abstract}
Empirical studies of polyglot repositories apply the same analyses to every language of a corpus, such as separating code from comments, computing
metrics, normalizing identifiers, and screening for duplicated code. Assembling them from specialized tools introduces differences in language coverage, measurement conventions, and workflows. We present \textsc{PyReprism}, an open-source Python library and command-line tool that separates language-specific lexical processing from reusable analyses: each language declares its lexical surface, which yields a lossless stream of typed tokens, and every analysis is implemented once on that stream. Across
\numLanguages{} built-in languages, language-specific declarations have a median length of \lexMedian{} lines, and no language overrides a shared analysis. Against a reference parser, comment and string classification reach F1 scores of \UFoneCmt{} and \UFoneStr{} on \UFiles{} files in \studyLangs{} languages that were never used to develop the tokenizer. In a study of a polyglot repository, one tool replaces three; with matched inputs and analysis operations, it takes \EQtimeP{}\,s instead of \EQtimeB{}\,s on one core.
We also characterize where lexical analysis falls short, in complexity estimation and clone screening.\\[2pt]
\textbf{Tool:} \url{https://github.com/unlv-evol/PyReprism}
\quad\textbf{Video:} \url{https://youtu.be/IXc-NP5t4GU}
\end{abstract}


\section{Introduction}
\label{sec:intro}

Software projects frequently combine multiple programming
languages~\cite{li2024multilingual}, so empirical studies of them must apply
the same analyses to every language of a corpus, such as measuring code
characteristics, separating comments from code, normalizing identifiers, and
detecting duplicated code~\cite{lopes2017dejavu}. Since no single tool
supports all of these analyses for all languages, researchers typically
combine tools that differ in language coverage, supported operations, and
measurement conventions. Such differences can change a study's outcome:
tools implementing the same metrics produce different values for the same
systems~\cite{lincke2008comparing}.

To put this challenge into perspective, consider \href{https://github.com/microsoft/pai}{\faGithub~OpenPAI}~\cite{openpai},
which we use as a running example throughout this paper. \href{https://github.com/microsoft/pai}{\faGithub~OpenPAI} combines
Bash, Python, Java, JavaScript, and TypeScript, so analyzing its
documentation, complexity, and duplication requires tools that differ in
both language coverage and measurement conventions. In our evaluation, we
analyzed it with \textsc{cloc}~\cite{cloc}, \textsc{Lizard}~\cite{lizard},
and \textsc{jscpd}~\cite{jscpd}, and the three tools did not even analyze
the same files: of the 718 files in these languages, \textsc{cloc} analyzed
667, \textsc{Lizard} 408, \textsc{jscpd} 735, because it counts JSX and TSX files twice. Per-language measurements from these tools therefore cover different sets of
files, which the researcher must identify, reconcile, and
report before any comparison. This motivates the shared lexical foundation
of \textsc{PyReprism}, in which analyses and their measurement policies are
implemented once and reused across languages, so that every analysis sees
the same files under the same conventions.

Researchers could instead build analyses on language infrastructure such as Pygments~\cite{pygments} or tree-sitter~\cite{brunsfeld2018treesitter},
but it provides tokens or syntax trees rather than metrics. Researchers
must still decide how token categories map to code and comment lines, what
counts as a decision point, and which tokens to rename, and then implement
and validate each analysis. Specialized tools thus offer complete analyses
for different sets of languages with their own conventions, while
infrastructure offers broad coverage but no analyses. To the best of our
knowledge, no existing tool lets researchers implement an analysis and its
measurement policy once and apply it to every language of a study. Yet many
of these analyses need only lexical information, including line accounting,
Halstead~\cite{halstead1977elements} and approximate
McCabe~\cite{mccabe1976complexity} metrics, identifier normalization,
$n$-gram naturalness~\cite{hindle2012naturalness}, and fingerprint-based
clone screening~\cite{schleimer2003winnowing}. This raises a practical
question: \emph{can a shared lexical representation support reusable
analyses across languages with sufficient accuracy for empirical research,
and what trade-offs does this introduce in precision, extensibility, and
performance?}

We present \textsc{PyReprism}, whose main contribution is a set of analyses
implemented once and reused for every supported language. Each language
declares only its lexical surface (file extensions, keywords, and comment and
string syntax), from which \textsc{PyReprism} produces a lossless stream of
typed tokens. All analyses operate on this stream, from line classification
and metrics to normalization, $n$-grams, fingerprints, and clone and diff
reports. Measurement policies, such as when a string counts as
documentation, are therefore defined once and applied uniformly to all
\numLanguages{} built-in languages. This does not remove methodological
choices, but it makes them explicit and consistent.

Our evaluation shows that no language overrides a shared analysis, that
comment and string classification reach F1 scores of 0.994 and 0.958 against
a reference parser on held-out files, and that in the OpenPAI study one tool
replaces three and runs a matched workload about three times faster, although
it is slower than \textsc{cloc} for counting lines alone. It also
characterizes where lexical analysis falls short: complexity estimates rank
files well but are not interchangeable with syntax-aware values, and lexical
clone screening finds copied and lightly edited fragments but few natural
clones.

\textsc{PyReprism} is an MIT-licensed Python package on PyPI
(\texttt{pip install PyReprism}), with source code on
GitHub~\cite{pyreprism_github} and documentation on Read the
Docs~\cite{pyreprism_docs}. A replication package~\cite{ogenrwot_2026_23612247} with all sampled files, scripts, runtimes, and tool versions accompanies this paper.

\section{Related Work}
\label{sec:related}

\emph{Specialized analysis tools.}
Several established tools support source-code analysis across programming
languages. \textsc{cloc}~\cite{cloc} and \textsc{scc}~\cite{scc} count source
lines, the latter also estimating complexity; \textsc{radon}~\cite{radon} and
\textsc{lizard}~\cite{lizard} compute software metrics; and
CCFinder~\cite{kamiya2002ccfinder}, SourcererCC~\cite{sajnani2016sourcerercc},
and \textsc{jscpd}~\cite{jscpd} identify duplicated code. These tools are
mature and together cover most analyses of a polyglot study, but each has its
own language coverage, file selection, and measurement conventions, and none
normalizes identifiers. Combining them therefore makes their integration part
of the study method. \textsc{PyReprism} does not outperform each tool at its
own task (e.g., \textsc{cloc} counts lines faster, and \textsc{PyReprism}'s
clone screening complements rather than replaces dedicated clone detectors).
Its benefit is running all analyses on the same files, with the same policies
and representation (Section~\ref{sec:workflow}).

\emph{Lexical infrastructure.}
Pygments~\cite{pygments} provides lexical analysis for more than 500
languages and is the closest underlying infrastructure to our approach, but
its token categories serve syntax highlighting rather than measurement.
Researchers using it must still map each token category to each
measurement, handle lexer-specific behavior, and implement and validate
every analysis themselves. \textsc{PyReprism} provides this missing layer:
32 token-based analysis methods and modules for metrics, fingerprints,
clones, $n$-grams, diffs, and batch processing, shared by all
\numLanguages{} languages. Because its measurement policies refer only to
token types (Section~\ref{sec:arch}), they apply unchanged whether tokens
come from the default dependency-free engine or from an optional Pygments
engine, although the engines classify some tokens differently
(Section~\ref{sec:correct}).

\emph{Syntax-aware infrastructure.}
\textsc{tree-sitter}~\cite{brunsfeld2018treesitter} and \textsc{srcML}~\cite{collard2013srcml}
expose structured representations of source code that enable analyses beyond
lexical tokens, but depend on per-language parsing support. Separating
language-specific front ends from analyses is a well-known design principle;
our contribution is a validated set of lexical analyses built on it, together
with an empirical characterization of when they suffice
(Section~\ref{sec:eval}). An optional \texttt{[ast]} extension provides
per-function measurements for languages with an available parser.


\section{Tool Description}
\label{sec:tool}
 
\subsection{Architecture and Token Stream}
\label{sec:arch}
\begin{figure}[t]
  \centering
  \includegraphics[width=\columnwidth]{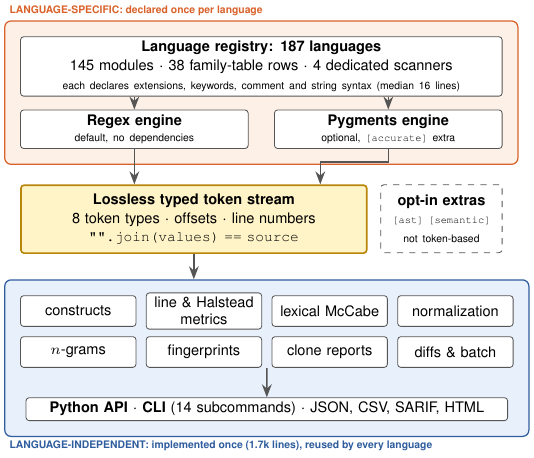}
  \caption{Architecture of \textsc{PyReprism}.}
  \label{fig:arch}
\end{figure}

Fig.~\ref{fig:arch} shows the architecture of \textsc{PyReprism}. In the
language-specific layer (top), a registry holds the lexical declaration of
each language (Section~\ref{sec:adding}), and a tokenization engine turns
source code into a token stream: either the dependency-free default engine,
which uses each language's regular expressions, or an optional Pygments
engine. In the language-independent layer (bottom), every analysis consumes
only that stream and is available through the Python API and a
\code{pyreprism} subcommand.
 
Each token has one of eight types (comment, string, number, keyword,
operator, identifier, whitespace, or other), a text, offsets, and a line number. The stream is \emph{lossless}: concatenating the token texts reproduces the input exactly. Transformations are filters that keep the order, content, and positions of the tokens they do not select, so their correctness depends on token classification (Section~\ref{sec:correct}). Policies that would otherwise differ between tools are defined once. A Python string that starts a line is a docstring and hence a comment (a lexical approximation of the first statement of a module, class, or function), as is a shebang line; preprocessor directives are code; a line is a comment line if it lies within a comment and contains no code, including blank lines inside block comments; and cyclomatic complexity is 1 plus the number of decision tokens in a file. Language-specific lexical rules, such
as nested comments, template literals, and regular-expression literals, stay in the language modules.
 
\subsection{Adding a Language}
\label{sec:adding}
A module declares a file extension, keywords, and a comment regular
expression whose named \code{comment} and \code{noncomment} groups separate comments from code and keep string literals whole (Listing~\ref{lst:ext}); string, number, and operator patterns have defaults that a module may override. Languages sharing comment syntax (C-style, hash, Lisp-style) are rows of a family table, and those requiring context, such as COBOL's column-sensitive comments or nested \code{/*~*/} comments, use small dedicated scanners. A registered language inherits every analysis, so the TOML module in Listing~\ref{lst:ext} immediately supports the full analysis layer.
 
\begin{lstlisting}[float=t,
caption={TOML declared as a module; all analyses are inherited.},
label=lst:ext]
@LanguageRegistry.register
class Toml(BaseLanguage):
    @classmethod
    def file_extension(cls): return ".toml"
    @classmethod
    def keywords(cls): return ["true", "false"]
    @classmethod
    def comment_regex(cls):
        return re.compile(r'(?P<comment>#.*?$)'
          r'|(?P<noncomment>"[^"]*"|.[^#"]*)',
          re.DOTALL | re.MULTILINE)
\end{lstlisting}
 
\subsection{Usage}
\label{sec:usage}
Listing~\ref{lst:usage} shows the Python API as used in the study of
Section~\ref{sec:workflow}. One \code{batch.map\_files} pass applies
\code{analyze} to every matching file on all cores, adds each file's metrics
to a per-language row, and saves a normalized, comment-free copy. Because
every language returns the same metric keys, the rows need no mapping
between tools. \code{clones.detect} then receives the paths of the same
files and reports Type-1/2 duplicated fragments and the share of duplicated
lines. The full script has 40 code lines. Each analysis is also a CLI
subcommand (Listing~\ref{lst:cli}): \code{scan} counts lines per language,
and \code{clones} reports duplicated fragments as text, JSON, SARIF,
Markdown, or HTML. With \code{--max-duplication}, \code{clones} exits with a
non-zero status, so a container image and a GitHub Action can use it as a
gate in continuous integration.
 
\begin{lstlisting}[float=t,caption={The workflow of Section~\ref{sec:workflow} through the Python API (abridged).},label=lst:usage]
from PyReprism import batch, clones
EXTS = ["*.py", "*.java", "*.js", "*.ts", "*.sh"]
 
def analyze(src, lang):            # runs once per file
    return lang.code_metrics(src), lang.normalize(src)
    
results = batch.map_files("pai", analyze,
                          include=EXTS, jobs=0)  # all cores
for path, lang, (m, norm) in results:
    rows[lang.__name__].add(m)     # same keys for all languages
    save(path, norm)               # ids renamed, no comments
files = [path for path, _, _ in results]  # same files as above
report = clones.detect(files, jobs=0)     # Type-1/2 fragments
\end{lstlisting}

\begin{lstlisting}[float=t,language={},columns=fixed,basewidth=0.5em,caption={Command-line analyses of OpenPAI (\code{pai}: \href{https://github.com/microsoft/pai}{\faGithub~microsoft/pai} at commit
\href{https://github.com/microsoft/pai/commit/43269e7}{\faCodeBranch~\code{43269e7}}; output abridged).},label=lst:cli]
% pyreprism scan pai --include '*.py' --include '*.java' \
      --include '*.js' --include '*.ts' --include '*.sh'
language    files  code  comment  blank
Bash          174  2018     3714   1370
Java          148 13510     3738   3248
JavaScript    124 12848     2445   1302
Python        156  8820     2651   3490
TypeScript     30  2774      480    346
% pyreprism clones pai/src --no-normalize --max-duplication 10
366 clones across 470 files; 24.25% duplicated lines
pyreprism: duplication 24.25% exceeds the 10.0% threshold
\end{lstlisting}

\section{Tool Evaluation}
\label{sec:eval}

We evaluate how much language-specific code the shared layer requires, how
accurately it classifies tokens, how it agrees with established tools, and
what it changes in an end-to-end study. We use \studyLangs{} languages
(Python, Java, C, C++, JavaScript, TypeScript, Go, C\#, Scala, Ruby, Bash, and
CSS), which cover the main comment conventions and have reference tools; nine
are among the ten most used on GitHub~\cite{octoverse2024}. From
\FrameFiles{} files in 501 cloned repositories and local open-source
installations, we drew 250 files per language (300\,B--200\,KB, round-robin
across projects, deduplicated by content): \corpusFiles{} files from
\SampleProjects{} projects. A disjoint draw of 120 files per language serves
the comparison with \textsc{cloc}. The replication package documents the
sampling and contains every file, script, and result.

\subsection{Extensibility}
\label{sec:ext}

We measure the language-specific code the shared layer requires, not the
effort to write or validate it. A recent release added 42 languages,
including COBOL and three nested-comment languages, in 392 lines of code and
140 lines of tests without changing any analysis. Of all \numLanguages{}
languages, 145 are modules, 38 are family-table rows, and 4 use a scanner.
Declarations have a median of \lexMedian{} lines (90th percentile:
\lexPninety{}; maximum: \lexMax{}), and none overrides any of the 32
token-based analysis methods of the base class, which together with the
shared modules form \sharedLoc{} lines of code. Short declarations are not
necessarily correct ones: several misclassified tokens until a reference
parser exposed them (Section~\ref{sec:correct}).

\subsection{Token Classification}
\label{sec:correct}

Losslessness shows that no character is lost, not that each is classified
correctly. We labeled every non-whitespace character as comment, string, or
code from \textsc{PyReprism}'s tokens and from a tree-sitter parse, a
reference parser rather than an infallible ground truth, which applies
\textsc{PyReprism}'s docstring policy; C macro bodies, which tree-sitter does
not parse, are not scored. We report character-level F1, micro-averaged per
language (Table~\ref{tab:correct}); excluding the \ParseErrFiles{} files with
parse errors changes scores by at most \CleanDelta{}.

Comparisons on two earlier development samples exposed misclassified strings
in eight languages, characters dropped by comment removal in 21 modules, and
preprocessor lines taken for comments by the Pygments engine; all were fixed
before the samples reported here (details in the replication package). On
the main sample, the default engine reaches pooled F1 scores of \FoneCmt{}
for comments and \FoneStr{} for strings (Pygments engine: \FoneCmtPyg{} and
\FoneStrPyg{}). Because this sample partly overlaps the development samples,
our key estimate comes from a held-out sample of \UFiles{} files from
\UProjects{} projects that excludes every development file and OpenPAI. On
it, the tokenizer reaches 0.994 and 0.958 (unweighted means over languages:
\UFoneCmtMean{} and \UFoneStrMean{}; per-language scores in Table~S1 of the
replication package). Ruby and Go are thinly represented in it
(\URubyFiles{} Ruby files from \URubyProjects{} projects; Go files from
\UGoProjects{} projects).

Concatenating the tokens reproduced every file exactly, but a high F1 does
not make every transformation safe. Judged by the reference,
\code{remove\_comments} deletes code in 33 of the \corpusFiles{} files
(1.1\%; column ``Del.'' of Table~\ref{tab:correct}) and in 7 of the
\UFiles{} held-out files (0.5\%), and leaves part of a comment in 24 files. Bash and
Ruby account for 24 of the 33 files, mostly through heredocs and string
interpolation, which need more context than a lexer has. Studies that use
transformed Bash or Ruby source should therefore validate it, for example by
parsing the output.

\begin{table}[t]
\caption{Token classification F1 against tree-sitter, files with incorrectly deleted code (Del.), and exact comment-line agreement with \textsc{cloc}. F1 and Del. use 250 files per language.}
\label{tab:correct}
\centering
\footnotesize
\setlength{\tabcolsep}{2.6pt}
\begin{tabular}{lcccc@{\hspace{0.9em}}lcccc}
\toprule
Lang. & Cmt. & Str. & Del. & \textsc{cloc} & Lang. & Cmt. & Str. & Del. & \textsc{cloc} \\
\midrule
\CorrectTableBody
\bottomrule
\end{tabular}
\end{table}

\subsection{Agreement with Established Tools}
\label{sec:agreement}

\emph{Line classification.} On the second draw (\ClocN{} files that
\textsc{cloc} recognized), comment-line counts match \textsc{cloc} exactly
for \ClocExact{} of the files (95\% CI \ClocCI{}; Table~\ref{tab:correct}),
with a median absolute difference of \ClocMAD{} lines and Spearman's
$\rho\geq\ClocRhoMin{}$ in every language. The low rates for Python and Bash
reflect policy: \textsc{cloc} counts blank lines inside docstrings and block
comments as blank, and shebang lines as code.

\begin{figure}[t]
\centering
\includegraphics[width=\columnwidth,height=5.2cm,keepaspectratio]{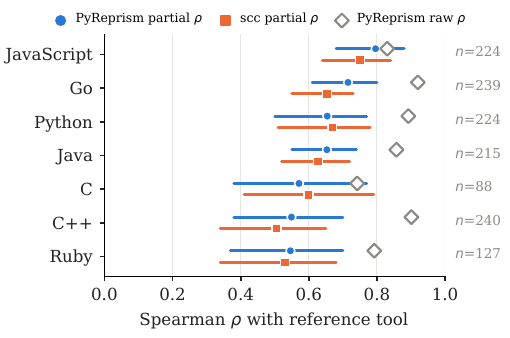}
\caption{Agreement of lexical cyclomatic complexity with
\textsc{radon} (Python) and \textsc{lizard} (other languages).
Dots and squares show size-adjusted Spearman $\rho$ for
\textsc{PyReprism} and \textsc{scc}, respectively;
diamonds show unadjusted $\rho$.
Error bars indicate bootstrap 95\% CIs.}
\label{fig:cc}
\end{figure}

\emph{Cyclomatic complexity.} The lexical estimate counts decision tokens per
file rather than walking a syntax tree. Since both it and the references grow
with file size, Fig.~\ref{fig:cc} reports size-adjusted Spearman
correlations. Against \textsc{radon} (Python) and \textsc{lizard} (others),
they range from \CCmin{} to \CCmax{}, and \textsc{scc}'s token-based estimate
is within \SCCdiff{} of \textsc{PyReprism} in every language. Low values
mostly reflect definitions: \textsc{lizard} adds~1 per function, a median
\JavaFunShare{} of a Java file's total. Against \textsc{lizard}'s decision
points alone, the partial $\rho$ is \DecRange{}, yet only \WithinTwo{} of
files are within $\pm 2$, so the values are not interchangeable. For
ranking, the lexical estimate recovers \TopOverlap{} of the files in
\textsc{lizard}'s top 10\% by decision points, against \TopSize{} for file
size alone, and beats size in every language. It thus suits shortlisting
complex files for inspection, not labeling individual files or replacing
syntax-aware values.

\subsection{An End-to-End Polyglot Study}
\label{sec:workflow}

We implemented the OpenPAI study of the introduction (\href{https://github.com/microsoft/pai}{\faGithub~microsoft/pai} at commit
\href{https://github.com/microsoft/pai/commit/43269e7}{\faCodeBranch~\code{43269e7}}) with \textsc{PyReprism} (Listing~\ref{lst:usage}) and with
\textsc{cloc}, \textsc{lizard}, and \textsc{jscpd} (Table~\ref{tab:workflow}).

\emph{Functional coverage.} Run with their defaults, both approaches report
per-language files, code lines, comment density, complexity, and duplicated
lines, and write comment-free copies. The toolchain needs three runtimes and
a mapping between three sets of language names, and its tools see different
subsets of the 718 files: \textsc{lizard} skips Bash and style sheets and
reports only files with functions (408), \textsc{cloc} skips files that
duplicate another (667), \textsc{jscpd} skips empty and very large files but
counts the 33 JSX and TSX files twice (735), and \textsc{PyReprism} skips
\texttt{build/} directories (691). None of these defaults is a defect, but
each changes the denominator of a per-language measurement and must be found
and reported when tools are combined. Moreover, \textsc{cloc}'s stripped
copies keep trailing comments, no tool renames identifiers, and definitions
change results: for Java, \textsc{PyReprism} finds \WFjavaTone{} duplicated
lines for identical token sequences and \WFjavaTtwo{} once identifiers are
renamed, against \WFjavaJscpd{} for \textsc{jscpd}. These default runs take
\WFtimeP{}\,s and \WFtimeB{}\,s on one core.

\emph{Equivalent workload.} To compare runtime on equal work, both approaches
processed the same \EQfiles{} files (the 691 minus empty files and three
one-line files that \textsc{jscpd} always skips, with \textsc{cloc}'s
duplicate check off and \textsc{jscpd}'s size limits raised) with four
matched operations: line counts, complexity for the six languages
\textsc{lizard} supports, Type-1 duplicated fragments, and comment-free
copies. The tools keep their own algorithms, so their outputs differ.
Table~\ref{tab:workflow} reports the median wall-clock time of ten runs,
start-up included, on a shared Apple M3 (IQR at most \EQiqr{}; all timings,
versions, and flags are in the replication package). This shows a faster
workflow, not faster individual operations: on the \corpusFiles{} corpus
files, \code{pyreprism scan} counts \TscanOne{} files per second on one core
against \Tcloc{} for \textsc{cloc} (\Tscan{} vs.\ \Tclocall{} on eight), and
the compiled \textsc{scc} reaches \Tscc{} but exposes no tokens for other
analyses. \textsc{PyReprism} is thus not the fastest choice for an isolated operation; its advantage in the evaluated workflow comes from serving several analyses through one representation.

\begin{table}[t]
\caption{OpenPAI workflow comparison. Runtime is the
median of ten runs, including start-up.}
\label{tab:workflow}
\centering
\footnotesize
\setlength{\tabcolsep}{3pt}
\begin{tabular}{lcc}
\toprule
 & \textsc{PyReprism} & \textsc{cloc}~/~\textsc{lizard}~/~\textsc{jscpd} \\
\midrule
\multicolumn{3}{l}{\emph{Functional coverage (each tool's defaults)}} \\
Tools / runtimes            & 1 / 1 & 3 / 3 (Perl, Python, Node.js) \\
Driver script (code lines)  & \WFlocP{} & \WFlocB{} \\
Files analyzed (of 718)     & 691 & 667 / 408$^{a}$ / 735$^{b}$ \\
Complexity for Bash         & yes & no \\
Identifier normalization    & yes & no \\
\midrule
\multicolumn{3}{l}{\emph{Equivalent workload (\EQfiles{} files, same operations)}} \\
Time, 1 core                & \EQtimeP{}\,s & \EQtimeB{}\,s \\
Time, 8 cores               & \EQtimePall{}\,s & \EQtimeBall{}\,s \\
\bottomrule
\end{tabular}\\[2pt]
{\scriptsize $^{a}$files with functions; $^{b}$JSX and TSX files counted twice.}
\end{table}

\subsection{Clone Screening}
\label{sec:clones}

Normalized fingerprints separate constructed Type-1, Type-2, and Type-3
clones of POJ-104 C/C++ solutions~\cite{mou2016poj104,lu2021codexglue} from
unrelated pairs (AUC \LadderAUC{}), a setting that favors token fingerprints
by design. On 1{,}000 natural Java clone pairs from
BigCloneBench~\cite{svajlenko2014bigclonebench}, however,
\textsc{PyReprism} detects only \BCBtpNear{} (recall \BCBrNear{}) and
\textsc{jscpd} \BCBtpJscpd{}, with no false positives on 1{,}000 non-clone
pairs, too few detections to establish precision. Lexical clone reports thus
screen for copied and lightly edited fragments; they are not a general clone
detector. The replication package contains both experiments in full.

\section{Limitations}

First, \textsc{PyReprism} relies on lexical information and cannot recover control flow, data flow, or bindings. Its complexity measures are therefore approximate, context-dependent constructs may be misclassified, and lexical clone screening misses weakly syntactic and semantic clones. Second, all \numLanguages{} languages are covered by regression tests, but comparison with a reference parser and established tools covers only \studyLangs{} languages; accuracy for the remaining languages is therefore not established. Third, a uniform measurement policy does not imply semantic equivalence across languages: constructs that document code or contribute to complexity can differ by language, which limits direct cross-language interpretation. Fourth, only the held-out sample estimates tokenizer accuracy on previously unseen files, and its coverage is limited for Ruby and Go. Finally, the evaluation has broader generalizability limits: the corpus is drawn from open-source software available on one machine rather than a defined project population, extensibility is measured by implementation size rather than developer effort, and the end-to-end workflow is evaluated on a single repository using scripts written by the authors.

\section{Conclusion}
\textsc{PyReprism} implements analyses once on a shared, lossless token
stream and applies them across languages. Our evaluation shows that this
reduces language-specific implementation, classifies comments and strings
close to a reference parser on unseen files, and simplifies a polyglot
workflow; it also shows that lexical measures are not interchangeable with
syntax-aware ones and that lexical clone screening misses semantic clones. A
shared lexical representation is thus a practical foundation for analyses of
heterogeneous repositories that do not depend on detailed program semantics.
Next, we validate the \code{[ast]} extension and integrate function-level complexity into the shared analyses.


\bibliographystyle{IEEEtran}
\bibliography{references}

\end{document}